\documentclass[12pt]{article}
\usepackage{epsf}
\usepackage{graphicx}
\usepackage{cite}
\usepackage{amsmath,amssymb}

\begin{document}

\noindent {\Large ANALYSIS OF SOURCE MOTIONS DERIVED FROM POSITION TIME SERIES}
\vspace*{0.7cm}

\noindent\hspace*{1.5cm} Z. MALKIN, E. POPOVA\\
\noindent\hspace*{1.5cm} Pulkovo Observatory\\
\noindent\hspace*{1.5cm} Pulkovskoe Ch., 65, St. Petersburg 196140, Russia\\
\noindent\hspace*{1.5cm} e-mail: malkin@gao.spb.ru\\

\vspace*{0.5cm}

\noindent {\large ABSTRACT.} %Your abstract of 150 words maximum
In this paper an attempt is made to extract
a systematic part from the source motions obtained from the position
time series provided by several IVS Analysis Centers in the framework
of the ICRF-2 project.  Our preliminary results show that the radio source
velocities and the parameters of the systematic part of the velocity field
differ substantially between the source position time series, which does
not allow us to get a reliable solution for the coefficients of spherical harmonics.

\vspace*{1cm}
\noindent {\large 1. INTRODUCTION}

\smallskip
Many radio sources observed during astrometric/geodetic VLBI sessions
show progressive variations in its position derived from single session
solutions.
Several physical effects can cause systematic apparent movement of celestial objects.
Hence investigation of the radio source apparent velocity field can help in investigations
in various fields, such as fundamental physics, cosmology, etc.
Several analysis strategies for computation of systematic part in the radio source velocities
can be used:
\begin{itemize}
\itemsep 0.5ex
\item[a)] estimate source position and velocities from global solution,
    then fit spherical harmonics to the velocities (Gwinn et al. 1997);
\item[b)] compute the coefficients of spherical harmonics as global parameters
    (MacMillan 2005; Titov 2008a, 2008b);
\item[c)] compute velocities from position time series,
    then fit spherical harmonics to the velocities.
\end{itemize}

In this paper, we will test the latter approach which, hopefully, can provide
a possibility for supplement comparisons and accuracy assessment.
\eject

%\vspace*{0.7cm}
\noindent {\large 2. DATA USED}

\smallskip
For this work we have used 26 source position time series computed at
9 VLBI analysis centers making use of 6 different software, which provides
a good opportunity for comparisons.

Table~\ref{tab:stat} presents data statistics.
In the ``All sessions'' columns, all submitted data are used having
at least two sessions (epochs) for the source, which allows us formally to compute
the velocity, even not of large scientific meaning.

For more rigorous comparison we also selected the data at common epochs for 17 series.
The common epochs were identified by the session name.
At this stage we did not use bkg000c series which seems to be just extension of
bkg000b with the same positions for the same sessions,
gsf000b series which seems to be only preliminary one,
aus series which does not contain session ID,
and several series with relatively small number of sources.
Statistics for thus selected data is shown in Table~\ref{tab:stat}
in the ``Common sessions'' columns.

\begin{table}
\begin{center}
\caption{Time series statistics}
\label{tab:stat}
\begin{tabular}{|l|l|cc|cc|}
\hline
~~Series & Software & \multicolumn{2}{|c|}{All sessions} &
         \multicolumn{2}{|c|}{Common sessions} \\
\cline{2-6}
        &               & Time span  & Nsou & Time span & Nsou \\
\hline
aus000a & OCCAM-LSC     & 1979--2007 & ~75 & ---        & --- \\
aus001a & OCCAM-LSC     & 1979--2007 & 578 & ---        & --- \\
aus002a & OCCAM-LSC     & 1979--2007 & 470 & ---        & --- \\
aus003a & OCCAM-LSC     & 1979--2007 & 503 & ---        & --- \\
bkg000b & Calc/Solve    & 1984--2007 & 615 & ---        & --- \\
bkg000c & Calc/Solve    & 1984--2007 & 794 & 1984--2007 & 463 \\
dgf000a & OCCAM-LS      & 1984--2007 & 425 & ---        & --- \\
dgf000b & OCCAM-LS      & 1984--2007 & 624 & 1984--2007 & 463 \\
dgf000c & OCCAM-LS      & 1984--2007 & 624 & 1984--2007 & 463 \\
dgf000d & OCCAM-LS      & 1984--2007 & 624 & 1984--2007 & 463 \\
dgf000e & OCCAM-LS      & 1984--2007 & 624 & 1984--2007 & 463 \\
dgf000f & OCCAM-LS      & 1984--2007 & 680 & 1984--2007 & 463 \\
dgf000g & OCCAM-LS      & 1984--2007 & 680 & 1984--2007 & 463 \\
gsf000b & Calc/Solve    & 1979--2005 & 721 & ---        & --- \\
gsf001a & Calc/Solve    & 1979--2007 & 742 & 1984--2007 & 463 \\
gsf002a & Calc/Solve    & 1979--2007 & 754 & 1984--2007 & 463 \\
iaa000b & QUASAR        & 1979--2007 & 540 & 1984--2007 & 463 \\
iaa000c & QUASAR        & 1979--2007 & 569 & 1984--2007 & 463 \\
mao000b & SteelBreeze   & 1980--2007 & 773 & 1984--2007 & 463 \\
opa000a & Calc/Solve    & 1984--2007 & 507 & ---        & --- \\
opa000b & Calc/Solve    & 1984--2007 & 645 & 1984--2007 & 463 \\
opa001a & Calc/Solve    & 1984--2007 & 519 & ---        & --- \\
opa002a & Calc/Solve    & 1984--2007 & 628 & 1984--2007 & 463 \\
sai000b & ARIADNA       & 1984--2007 & 640 & 1984--2007 & 463 \\
usn000d & Calc/Solve    & 1979--2007 & 728 & 1984--2007 & 463 \\
usn001a & Calc/Solve    & 1979--2007 & 728 & 1984--2007 & 463 \\
\hline
\end{tabular}
\end{center}
\end{table}

\vspace*{0.7cm}
\noindent {\large 3. COMPARISON OF VELOCITIES}

\smallskip
The source velocities were computed as weighted linear drift of the submitted
source positions with weights inversely proportional to the reported
variances of source positions.
Since some time series contain positions with unlikely small errors
(down to 1 $\mu$as in the iaa series), which leads to problems with computing
the velocity as the weighted trend, it was decided to use a minimal
error value of 20 $\mu$as, i.e. all errors less then this value were
replaced by 20 $\mu$as.
No series except iaa were substantially affected by this procedure.

Our programs for computation of source velocities and spherical harmonics has several optional
parameters for data selection: the first epoch, minimum number of sessions, minimum data span,
maximum error in velocity.
The results presented in this section were computed for all the data presented in the original
series having at least 5 sessions and 3-year time span.

In Table~\ref{tab:verr}, the results of computation of the median error in velocity
computed for are presented.
This values may be an index of the scatter of position time series.
One can see that some time series are much more noisy than others.

\begin{table}
\begin{center}
\caption{Median errors in velocities. Unit: $\mu$as/yr}
\label{tab:verr}
\tabcolsep 10pt
\begin{tabular}{|l|cc|cc|}
\hline
~~Series & \multicolumn{2}{|c|}{All sessions} &
         \multicolumn{2}{|c|}{Common sessions} \\
\cline{2-5}
& $V_{\alpha}\cos\delta$ & $V_\delta$ & $V_{\alpha}\cos\delta$ & $V_\delta$ \\
\hline
aus000a & ~7 & 11 & -- & -- \\
aus001a & 19 & 28 & -- & -- \\
aus002a & 18 & 26 & -- & -- \\
aus003a & 18 & 29 & -- & -- \\
bkg000c & 14 & 18 & 17 & 21 \\
dgf000a & 19 & 25 & -- & -- \\
dgf000b & 15 & 18 & 18 & 21 \\
dgf000c & 15 & 21 & 19 & 26 \\
dgf000d & 15 & 19 & 19 & 23 \\
dgf000e & 15 & 19 & 19 & 22 \\
dgf000f & 16 & 23 & 19 & 26 \\
dgf000g & 16 & 19 & 19 & 23 \\
gsf001a & 13 & 17 & 15 & 19 \\
gsf002a & 13 & 17 & 17 & 20 \\
iaa000b & 15 & 22 & 18 & 22 \\
iaa000c & 16 & 22 & 20 & 23 \\
mao000b & 23 & 31 & 25 & 34 \\
opa000a & 15 & 19 & -- & -- \\
opa000b & 16 & 23 & 19 & 28 \\
opa001a & 15 & 18 & -- & -- \\
opa002a & 17 & 19 & 16 & 20 \\
sai000b & 25 & 37 & 30 & 45 \\
usn000d & 13 & 18 & 17 & 20 \\
usn001a & 21 & 29 & 25 & 34 \\
\hline
\end{tabular}
\end{center}
\end{table}

\begin{figure}
\begin{center}
\includegraphics[scale=0.5]{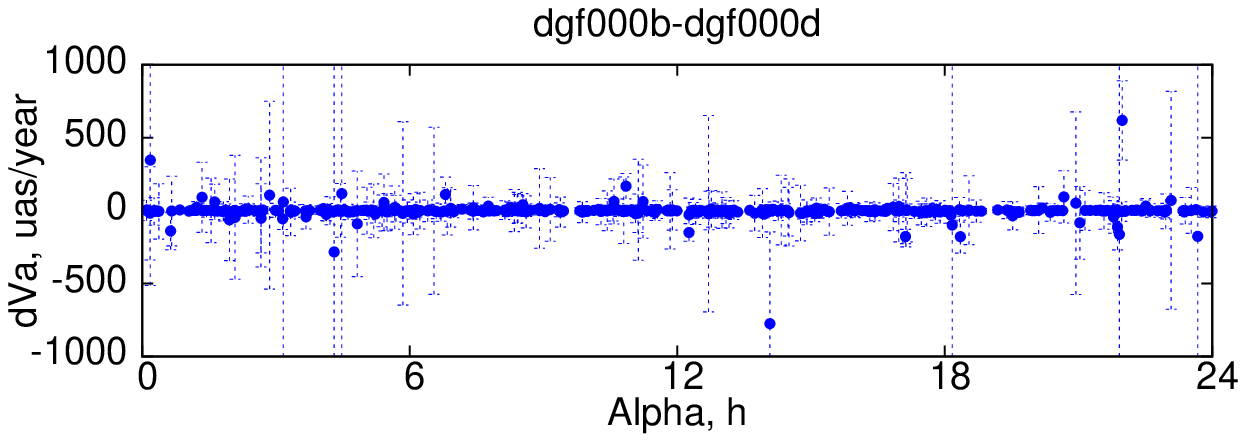}\hskip 2em
\includegraphics[scale=0.5]{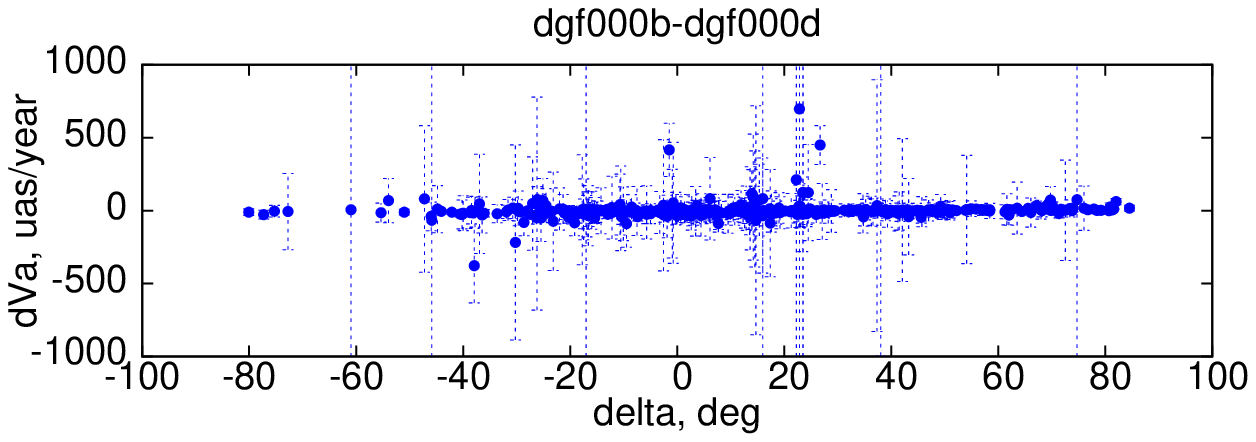}

\includegraphics[scale=0.5]{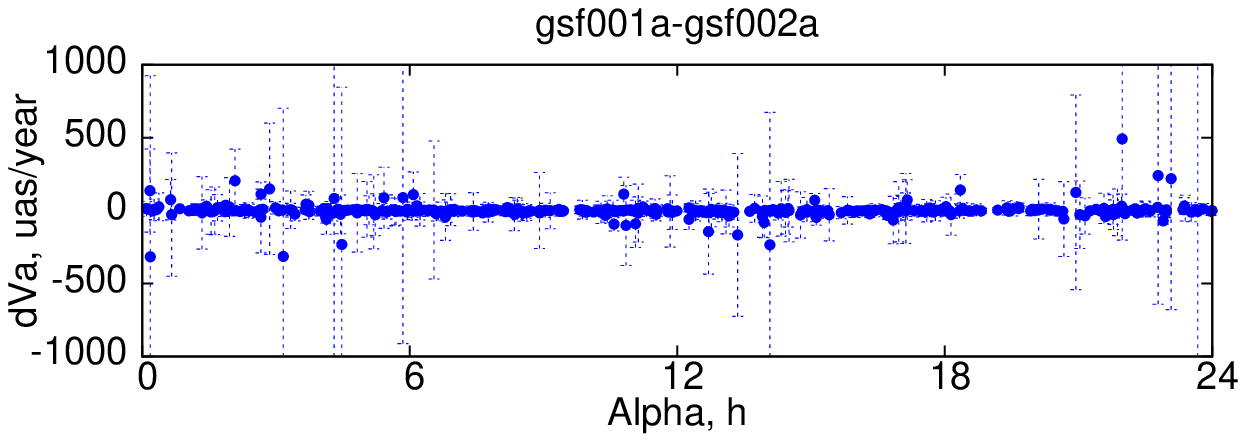}\hskip 2em
\includegraphics[scale=0.5]{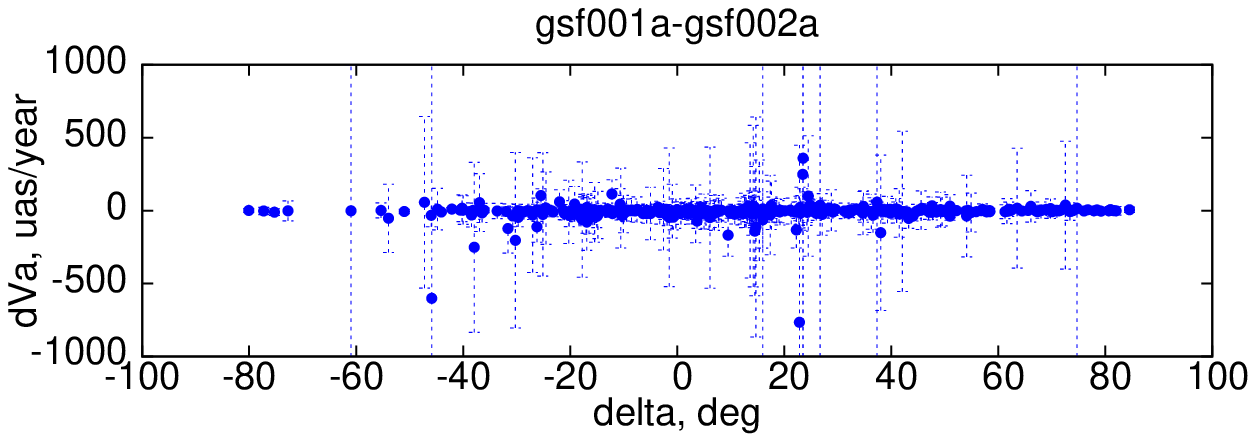}

\includegraphics[scale=0.5]{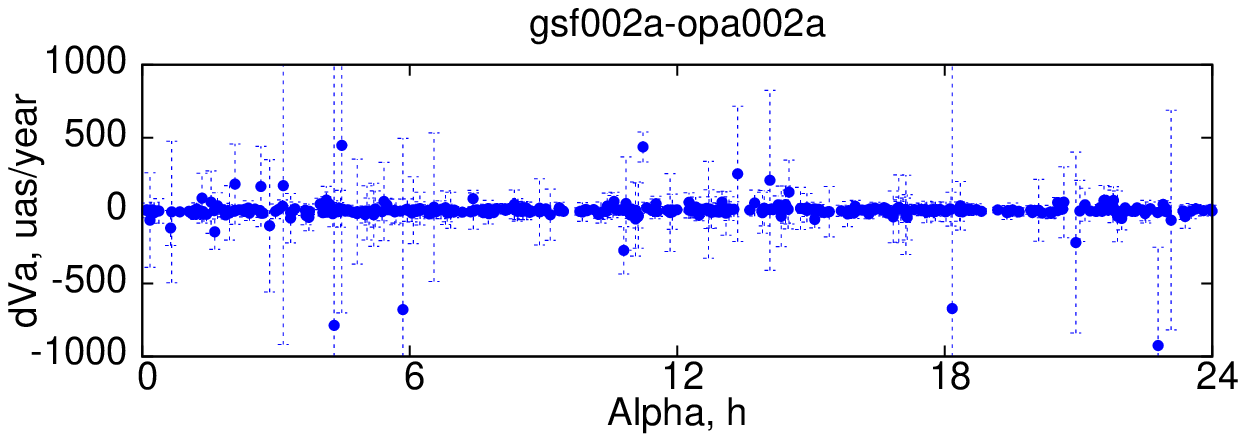}\hskip 2em
\includegraphics[scale=0.5]{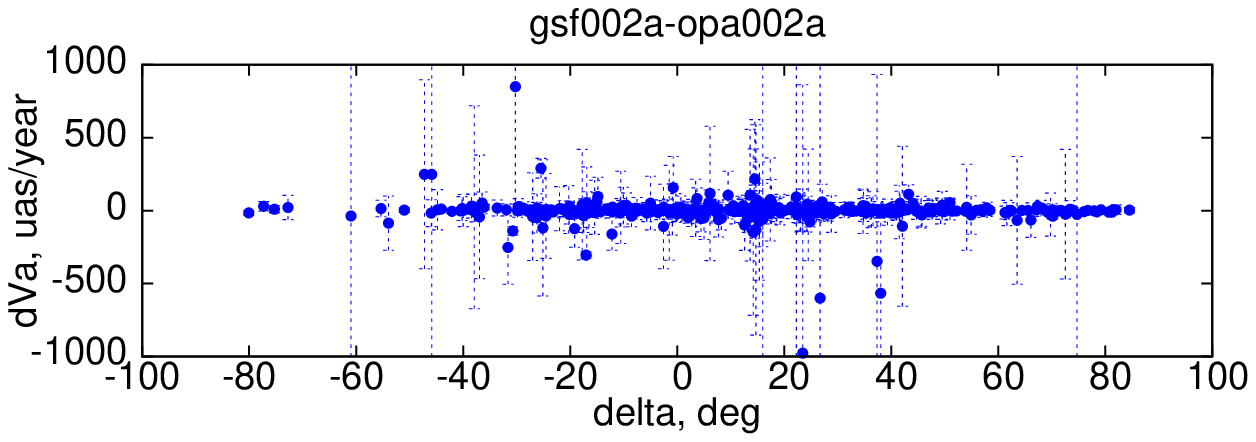}

\includegraphics[scale=0.5]{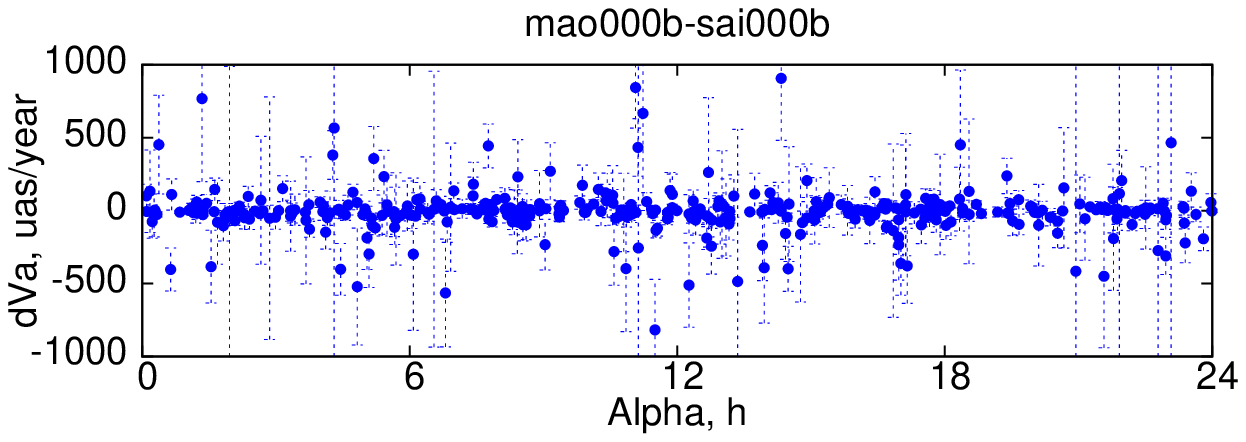}\hskip 2em
\includegraphics[scale=0.5]{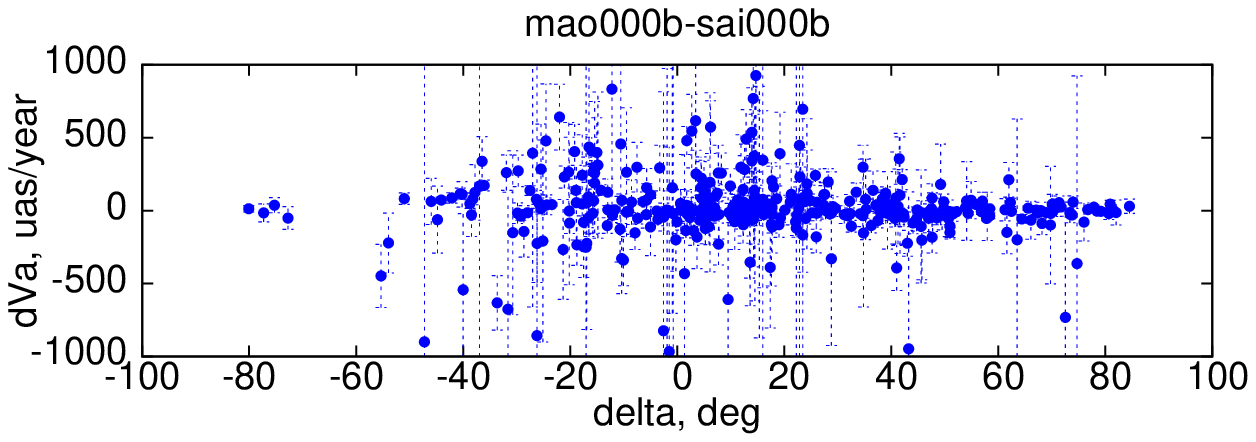}

\includegraphics[scale=0.5]{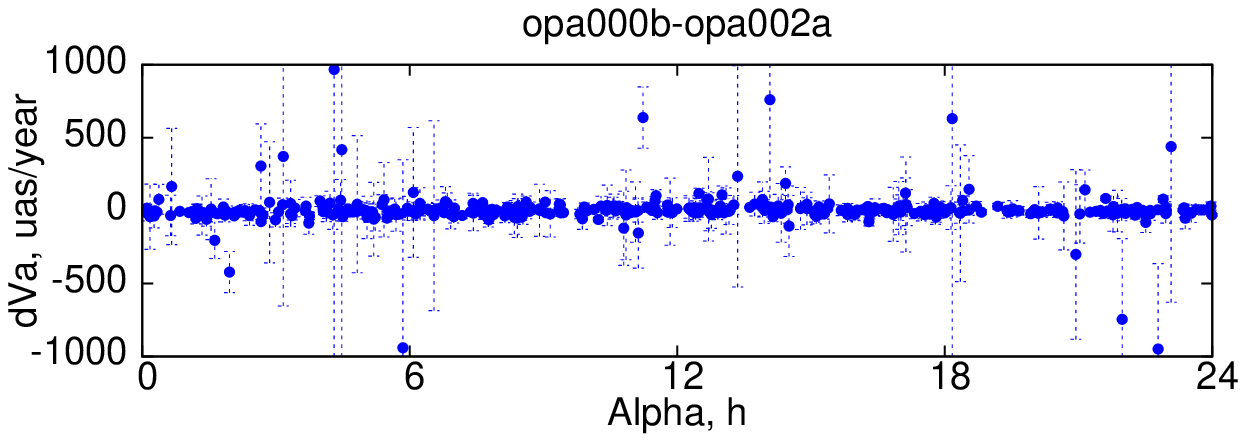}\hskip 2em
\includegraphics[scale=0.5]{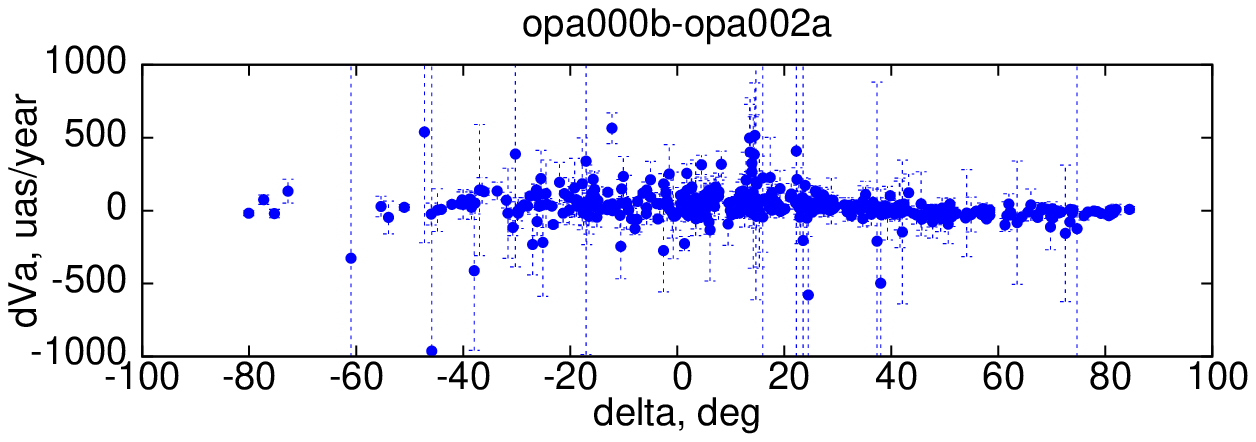}

\includegraphics[scale=0.5]{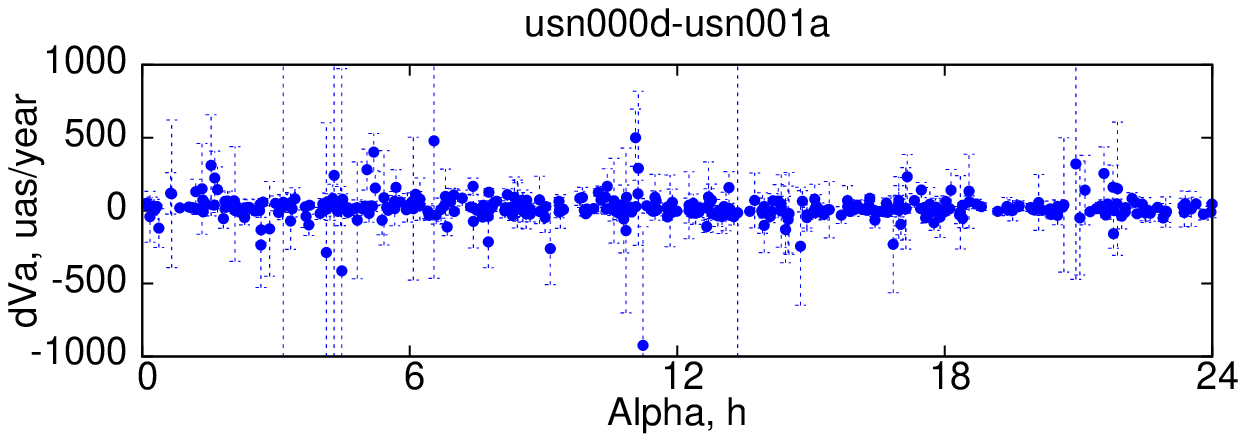}\hskip 2em
\includegraphics[scale=0.5]{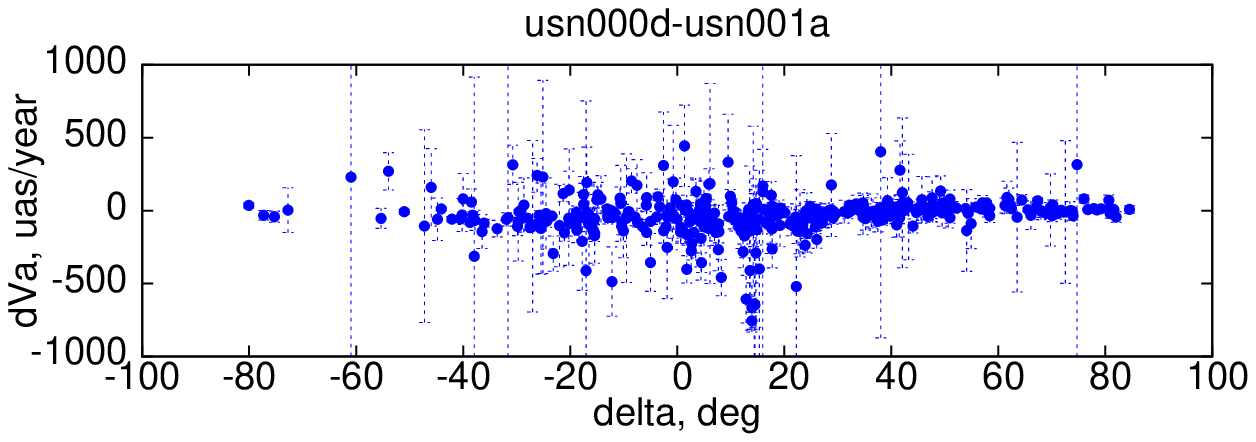}
\caption{Comparison of source velocities obtained from different time series}
\label{fig:vel_diff}
\end{center}
\end{figure}

Comparison of velocities obtained from different time series for
the same source show sometimes rather large discrepancies.
In Fig~\ref{fig:vel_diff} some examples are given
(only data at common epochs were used for rigorous
comparison).

%\vspace*{0.7cm}
\noindent {\large 4. TEST OF HARMONICS}

\smallskip

For test purpose, we compute the  coefficients of two spherical harmonics
$\Delta H_{12}$ and $\Delta H_3$ (Titov 2008a) using the following formulas:
\begin{eqnarray*}
\mu_\alpha & = & -\Delta H_{12} \sin 2\alpha \,, \\[1em]
\mu_\delta & = & -\frac{\Delta H_{12}}{4} \cos 2\alpha \sin 2\delta + \frac{\Delta H_3}{2} \sin 2\delta.
\end{eqnarray*}

We tried several options for data selection which produce different results; however,
the difference is usually not so large for reasonable values of the selection criteria given above.
In Table~\ref{tab:h123}, the results are presented computed for all the data presented in the original
series having at least 5 sessions and 3-year time span.
It can be noted that using more strict criteria, such as minimum 10 sessions and 10 years of
observations gives statistically similar result, with smaller value of the formal
error in the harmonics coefficients when more observations are used.
In the last row of the table the results are presented corresponding to the
cumulative solution including all the velocity estimates merged all the input time series.

\begin{table}
\begin{center}
\caption{Results of computation of $\Delta H_{12}$ and $\Delta H_3$. Unit: $\mu$as}
\label{tab:h123}
\begin{tabular}{|l|cr@{$\pm$}lr@{$\pm$}l|cr@{$\pm$}lr@{$\pm$}l|}
\hline
~~Series & \multicolumn{5}{|c|}{All sessions} &
         \multicolumn{5}{|c|}{Common sessions} \\
\cline{2-11}
       & Nsou & \multicolumn{2}{c}{$\Delta H_{12}$} & \multicolumn{2}{c|}{$\Delta H_3$} &
         Nsou & \multicolumn{2}{c}{$\Delta H_{12}$} & \multicolumn{2}{c|}{$\Delta H_3$} \\
\hline
aus000a &    71  & -8.76 & 3.15 &  1.02 & 2.33 &  --- & \multicolumn{2}{c}{---} & \multicolumn{2}{c|}{---} \\
aus001a &   343  & -4.53 & 1.08 & -2.13 & 0.87 &  --- & \multicolumn{2}{c}{---} & \multicolumn{2}{c|}{---} \\
aus002a &   308  & -0.34 & 1.22 & -0.47 & 0.99 &  --- & \multicolumn{2}{c}{---} & \multicolumn{2}{c|}{---} \\
aus003a &   322  & -3.80 & 1.12 & -2.47 & 0.90 &  --- & \multicolumn{2}{c}{---} & \multicolumn{2}{c|}{---} \\
bkg000c &   537  & -1.01 & 0.83 &  0.85 & 0.74 &  350 &  -0.27 & 0.91 & -1.27 & 0.83 \\
dgf000a &   277  & -3.84 & 1.44 &  4.88 & 1.49 &  --- & \multicolumn{2}{c}{---} & \multicolumn{2}{c|}{---} \\
dgf000b &   476  & -1.66 & 0.73 &  1.12 & 0.70 &  350 &  -1.06 & 0.74 & -0.73 & 0.71 \\
dgf000c &   476  & -0.27 & 0.90 &  1.23 & 0.76 &  350 &   0.88 & 0.93 & -1.08 & 0.80 \\
dgf000d &   476  & -1.94 & 0.75 &  1.14 & 0.72 &  350 &  -1.47 & 0.77 & -0.71 & 0.73 \\
dgf000e &   476  & -2.08 & 0.73 &  1.08 & 0.71 &  350 &  -1.71 & 0.77 & -0.57 & 0.73 \\
dgf000f &   531  & -0.29 & 0.86 &  1.27 & 0.71 &  350 &   0.85 & 0.97 & -0.88 & 0.81 \\
dgf000g &   531  & -2.09 & 0.71 &  1.18 & 0.67 &  350 &  -1.83 & 0.80 & -0.43 & 0.75 \\
gsf001a &   582  & -0.62 & 0.70 & -0.09 & 0.64 &  350 &  -0.77 & 0.86 & -0.76 & 0.78 \\
gsf002a &   592  & -0.39 & 0.65 &  0.64 & 0.59 &  350 &   0.50 & 0.83 & -1.46 & 0.76 \\
iaa000b &   458  &  1.23 & 1.37 &  0.80 & 1.16 &  350 &   3.49 & 1.21 &  1.70 & 1.08 \\
iaa000c &   481  &  1.15 & 1.34 &  2.16 & 1.15 &  350 &   2.75 & 1.24 &  3.44 & 1.17 \\
mao000b &   555  &  0.05 & 1.05 &  1.01 & 0.86 &  350 &   0.14 & 1.26 & -0.35 & 1.07 \\
opa000a &   384  &  0.11 & 1.10 & -0.46 & 0.95 &  --- & \multicolumn{2}{c}{---} & \multicolumn{2}{c|}{---} \\
opa000b &   510  & -6.14 & 1.09 &  0.46 & 0.88 &  350 & -10.55 & 1.52 & -1.01 & 1.21 \\
opa001a &   392  &  0.20 & 0.99 & -0.37 & 0.87 &  --- & \multicolumn{2}{c}{---} & \multicolumn{2}{c|}{---} \\
opa002a &   511  &  0.66 & 0.87 & -0.53 & 0.77 &  350 &  -0.21 & 0.94 &  0.15 & 0.85 \\
sai000b &   501  & -1.96 & 1.18 &  1.79 & 0.95 &  350 &   0.27 & 1.37 & -1.39 & 1.14 \\
usn000d &   572  & -0.70 & 0.78 &  0.18 & 0.70 &  350 &  -0.25 & 0.90 & -1.44 & 0.81 \\
usn001a &   572  & -5.94 & 1.22 &  1.53 & 1.02 &  350 &  -6.57 & 1.67 & -0.28 & 1.38 \\
\hline
All data& 11477  & -1.24 & 0.19 &  0.62 & 0.17 & 5950 &  -0.54 & 0.23 & -0.51 & 0.21    \\
\hline
\end{tabular}
\end{center}
\end{table}

\vspace*{0.7cm}
\noindent {\large 5. CONCLUDING REMARKS}

\smallskip
Although most results presented in Table~\ref{tab:h123} are formally
statistically reliable, they differ substantially between input time series,
and also between various sets of data selected.
This fact, along with results of velocity comparison
may indicate that source position time series should be
used with care for analysis of the fine effects in the source proper motions.

Further study is needed to investigate a possibility to
use combined or cumulative solution as the most reliable estimate of
spherical harmonics.
In particular, careful selection of input series should be performed.
For instance, in our cumulative solution dgf data are clearly overweighted
due to 6 series used, often with very similar position estimates.
On the other hand, it seems to be inappropriate to use only one series
from one analysis center because some centers compute two and more
series using quite different approaches, and this would be important
to compare all of them, because there is no indisputable proof in favor
of only one approach.

\clearpage
\noindent {\large 6. REFERENCES}
% Please type the reference as follows
% Name Initial, year, "title", journal, vol. , pp. x-x.
%
% Examples:
%
% Author1, N., Author2, N., 2000, ``Title of the paper'',
% \aa 111, pp. 111--222.
%
% Author2, N., Author3, N., 2003, ``Title of the paper'',
% \jgr (Solid Earth), 111(B5), doi: 10.1000/2002JB001111.
%
% PLEASE DO NOT USE ANY SPECIAL FONTS
% (no italics, no boldface, etc.)

{

\leftskip=5mm
\parindent=-5mm

\smallskip

Gwinn, C. R., et. al., 1997, "Quasar proper motions and low-frequency
gravitational waves", ApJ 485, pp. 87-91.

Macmillan, D.S., 2005,
``Quasar Apparent Proper Motion Observed by Geodetic VLBI Networks'',
In: Future Directions in High Resolution Astronomy: The 10th Anniversary
of the VLBA, ASP Conference Proceedings, V.~340. Eds. J.~Romney, M.~Reid,
San Francisco: Astronomical Society of the Pacific, 2005, pp. 477--481.

Titov. O., 2008a, ``Proper motions of reference radio sources'',
In: Proc. Journ\'ees Syst\`emes de R\'ef\'erence Spatio-temporels 2007,
Meudon, France, 17-19 Sep 2007, Ed. N.~Capitaine, pp.~16--19.

Titov, O., 2008b, ``Systematic effects in proper motion of radio sources'',
arXiv:0805.1099.

}

\end{document}